# Modular Multirotary Joints SPS Concept—Challenges and Design Considerations

XINBIN HOU
ZHENGAI CHENG
China Academy of Space Technology, Beijing, China

XIN WANG, Member, IEEE
Chongqing University, Chongqing, China

CHANGJUN LIU, Senior Member, IEEE
Sichuan University, Chengdu, China

As a major space-based energy infrastructure in response to the challenge of climate change, space solar power (SSP) has attracted extensive international attention for more than 55 years. Recently, various SSP concepts have been proposed, aiming to efficiently harvest gigawatts of solar power over Earth's geostationary orbit and transmit it to the Earth wirelessly. Based on the multirotary joints solar power satellite (MR-SPS) concept proposed in 2014, this article presents an updated modular MR-SPS (MMR-SPS) concept, together with the challenges and design considerations. The primary objective is to address the challenges associated with long-distance dc electric power transmission and exceptionally high-precision microwave power transmission. The MMR-SPS employs multiple modules, each further including a solar subpanel and an antenna subpanel. The solar subpanel and antenna subpanel in each module are connected with each other via conductive rotary joints with a much shorter distance compared with that in the MR-SPS concept. As a result, lower voltage in electric power transmission is allowed, which significantly reduces the complexity of the system and increases its reliability. Retroreflective beamforming is employed to ensure the accurate steering of the microwave power beam to the ground's receiving aperture. For the first time, this article analyzes and validates the performance of using retroreflective beamforming to compensate for antenna position deviations through a numerical calculation model that includes near-field effects. Additionally, this article outlines the transportation and assembly methods, along with the considerations of the full life-cycle cost for an MMR-SPS.

Manuscript received 13 March 2024; revised 5 June 2024 and 30 July 2024; accepted 5 August 2024. Date of publication 12 August 2024; date of current version 11 February 2025.

DOI. No. 10.1109/TAES.2024.3441570

Refereeing of this contribution was handled by Rick Meyer.

This work was supported in part by Civil Aerospace Technology Research Project under Grant D010103, and in part by the National Natural Science Foundation of China under Grant U22A2015 and Grant 61871220.

Authors' addresses: Xinbin Hou and Zhengai Cheng are with the China Academy of Space Technology, Beijing 100094, China, E-mail: (houxinbin525@163.com), (alicechengqlab@163.com); Xin Wang is with Chongqing University, Chongqing 400044, China, Email: (wang.x@cqu.edu.cn); Changjun Liu is with Sichuan University, Chengdu 610065, China, E-mail: (cjliu@ieee.org). (Corresponding author: Xin Wang.)

## I. INTRODUCTION

The concept of space power satellite (SPS), also known as space solar power (SSP), was proposed in 1968 [1]. It aims to harvest gigawatts of solar power from satellites over Earth's geostationary orbit (GEO) and then deliver the power to Earth wirelessly. Unlike terrestrial solar power, SSP does not suffer from nighttime limitations and weather disruptions, offering a new source of sustainable clean energy with continuous availability. Hence, SPS has been considered as one of the most promising energy projects addressing the challenges of energy crisis and climate change. Various development plans and research activities related to SPS are underway in the United States, Japan, China, South Korea, and several European countries [2], [3], [4], [5], [6], [7], [8].

The basic concept of SPS is illustrated in Fig. 1. An SPS typically includes a solar panel and an antenna panel. The solar power is harvested and converted to electric power in direct current (dc) by the solar cells on the solar panel. The dc electric power is delivered to the antenna panel via electric cables. The hardware integrated within the antenna panel converts the dc electric power to microwave power and controls the beamforming of the microwave power toward Earth. Given the dependency of atmospheric transparency on frequencies, 5.8 GHz is considered an excellent candidate for the wireless power transmission [9]. This is also within the frequency range recommended for wireless power transmission via radio frequency beam by the International Telecommunication Union Radiocommunication sector [10]. To ensure the efficient operation of the SPS, it is preferable to keep the solar panel facing the Sun and the antenna panel facing the Earth. Specifically, with the SPS located over the GEO, the antenna panel facing Earth can be kept stationary, while the solar panel must be dynamically adjusted toward the Sun. Therefore, managing the mutual position relationship between the solar panel and the transmitting antenna panel has become one of the core issues in the SPS design. Another technical issue in SPS design is minimizing the power loss during electric power transfer from the solar panel to the antenna panel. Space environmental factors prohibit the use of high-voltage techniques commonly employed in terrestrial electric power systems, resulting in significant power dissipation during long-distance power transmission [11], [12]. Dozens of SPS concepts have been proposed to tackle these challenges.

Fig. 2 illustrates some of the recently proposed concepts (a comprehensive review of all SPS concepts is beyond this article's scope). The MR-SPS [13] and the K-SSPS [7] concepts use conductive rotary joints to maintain Sun pointing of the solar panel and Earth pointing of the antenna panel. These designs enable a consistent and uninterrupted generation of power, but require high-power conductive rotary joints and long cables to deliver electric power from the solar panel to antennas. The SPS-ALPHA [14], SSPS-OMEGA [15], and CASSIOPeiA SPS concepts [16] utilize specialized concentrator to redirect sunlight toward the solar panel, eliminating the need for high-power conductive rotary joints and lengthy electric cables. However, challenges arise in achieving precise control and effective heat dissipation for the concentrator. The tethered-SPS [17] and the Caltech SPS concepts [18], [19] feature a fixed position of the solar panel relative to the antenna, resulting in varying angles between the solar panel and the Sun during orbit, which causes significant fluctuations in collected solar energy during the orbital period. Effectively managing these fluctuations constitutes a main challenge in these configurations.

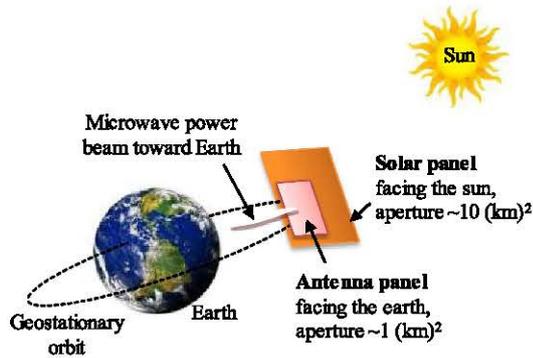

Fig. 1. Illustration of the basic concept of an SPS (not to scale).

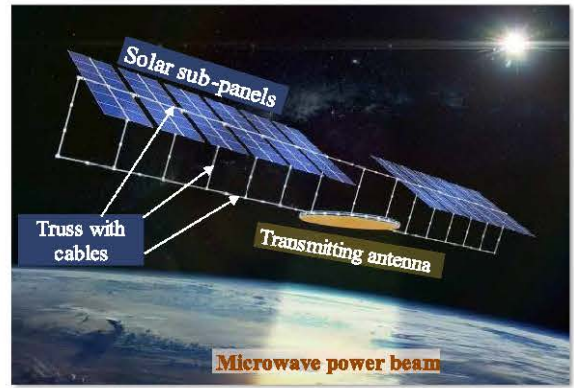

(a)

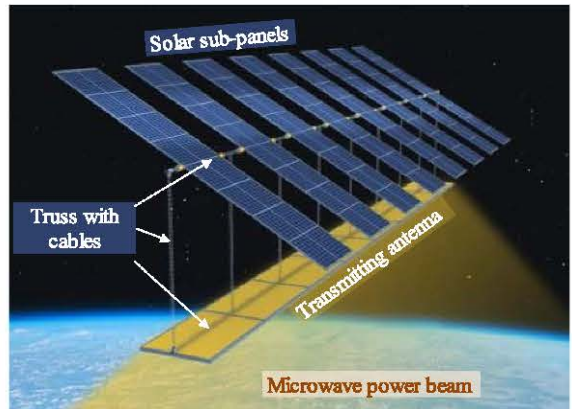

(b)

Fig. 3. Architecture of the (a) MR-SPS concept compared with the (b) proposed MMR-SPS concept.

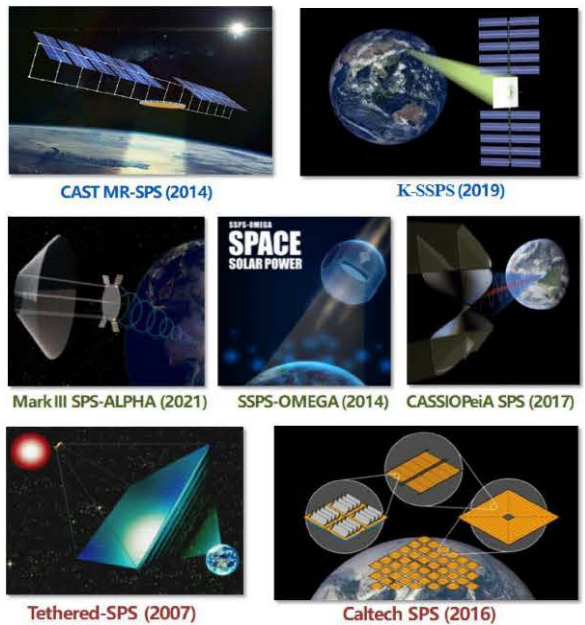

Fig. 2. Illustrations of typical SPS concepts.

Based on the multirotary joints solar power satellite (MR-SPS) concept proposed in 2014, this article presents an updated modular MR-SPS (MMR-SPS) concept, which aims to address the challenges associated with long-distance dc electric power transmission [20]. As illustrated in Fig. 3(a), the solar panel of the MR-SPS is divided into multiple subpanels. Multiple rotary joints along the upper truss structure are employed to orient the solar subpanels toward the Sun, which significantly reduces the power handling requirement for each conductive rotary joint. The issue of a single point of failure in the conductive rotary joint is also avoided. However, the difficulty of long-distance dc electric power transmission remains a substantial technical challenge. This is because the antenna panel is in the form of a circular disk aggregated at the center of the satellite, which necessitates the transmission of electric power from the solar subpanels via cables spanning several kilometers. Due to the limitation on the voltage level that can be used in space, the power loss over such distance is significantly high. To address this challenge, the MMR-SPS employs a linear rectangular antenna panel, as illustrated in Fig. 3(b). The rectangular antenna panel comprises multiple subpanels distributed along the lower truss structure beneath the solar subpanels. Each solar subpanel combines with one antenna subpanel to form an independent solar power generation and microwave power transmission (SPG-MPT) module. The power generated by each solar subpanel is transmitted to the corresponding antenna subpanel through much shorter cables. This design significantly mitigates the challenges associated with space power transmission and management (PTM) while also simplifying the SPS transportation and assembly.

Another technical challenge addressed in this article is the demand for extremely high accuracy in beamforming and steering in microwave power transmission (MPT) from the antenna panel on the satellite to the receiving aperture on Earth. The modular configuration of the MMR-SPS leads to a unique long-strip-shaped antenna array layout, presenting a distinct challenge in beamforming control design. Particularly concerned in this article is the effect of antenna position deviations and their compensation via retroreflective beamforming.

The rest of this article is organized as follows. The system architecture of the proposed MMR-SPS concept is explored in Section II. The technique of achieving exceptionally high-precision MPT based on retroreflective beamforming under the influence of antenna position deviations is deliberated in Section III. Section IV presents a proposed in-orbit transportation and assembly scheme, followed by an analysis of the life-cycle cost in Section V. Finally, Section VI concludes this article.

## II. SYSTEM ARCHITECTURE OF THE PROPOSED MMR-SPS

As an extremely complex engineering system, an SPS architecture includes a large number of technical elements, and it is impossible for this section to cover all of them. Rather, three major elements of the MMR-SPS concept, including the SPG-MPT module design, electric PTM, and the MPT, are portrayed in the following sections. Several important technical elements not covered by this article (such as structural framework, attitude and orbit control, thermal management, and information and system operation management) will be investigated separately in other articles.

The following discussions are based on the predictive energy efficiency chain of a 1 GW MMR-SPS, as shown in Table I. The table outlines the target energy efficiencies due to various factors considered in the proposed MMR-SPS concept. Some of them are estimated from the published research. For instance, the best research-cell efficiency chart in [21] shows that certain photovoltaic technologies, such as multijunction cells, have achieved efficiencies higher than 40%. The values of other efficiencies are based on the estimations provided in [22] and [23], which are deemed to be reasonable targets, given the expected progress in the related fields over the coming decades. Some of the factors are unique to the MMR-SPS concept, including the Sun-pointing error, angle of sunlight, and space environment effects [24]. The overall system efficiency is the cumulative energy efficiency in Table I, i.e., 12.3%. In order to generate 1 GW dc electrical power output on Earth, the solar power incident upon the solar panel of the SPS should reach around 8.2 GW. Given the solar constant to be 1.367 kW/m$^2$ [25], the area of the solar panel should be at least (8.2 GW)/(1.367 kW/m$^2$) = $6 \times 10^6$ m$^2$. This leads to the physical dimensions of the solar panel adopted in the MMR-SPS as 600 m $\times$ 10 000 m.

TABLE I
Predictive Energy Efficiency Chain

| Factor | Factor efficiency | Cumulative efficiency |
|---|---|---|
| Efficiency in solar power collection and conversion: 0.29 | | |
| Solar Cell | 0.40 | 0.4 |
| Error of Sun pointing | 0.99 | 0.396 |
| Solar Array Filling Factor | 0.85 | 0.336 |
| Angle of Sunlight | 0.958 | 0.322 |
| Space Environment Effects | 0.90 | 0.290 |
| Efficiency in dc PTM: 0.89 | | |
| DC–DC in solar array | 0.97 | 0.281 |
| Rotary joints and cables | 0.95 | 0.267 |
| DC–DC in Antennas | 0.97 | 0.259 |
| Service Devices Consumption | 0.999 | 0.258 |
| Efficiency in dc–microwave conversion and radiation: 0.76 | | |
| DC–Microwave | 0.80 | 0.207 |
| Microwave radiation | 0.95 | 0.197 |
| Efficiency in microwave transmission in atmosphere: 0.95 | | |
| Microwave transmission | 0.95 | 0.187 |
| Efficiency in microwave power collection and conversion: 0.69 | | |
| BCE | 0.9 | 0.168 |
| Receiving antenna efficiency | 0.9 | 0.151 |
| Microwave–dc | 0.85 | 0.129 |
| Efficiency in DC power combination and conversion: 0.96 | | |
| Power Collection | 0.98 | 0.126 |
| DC–DC | 0.98 | 0.123 |

### A. SPG-MPT Module Design

Fig. 4 depicts the specific configuration of a typical 1 GW MMR-SPS architecture. It comprises 50 SPG-MPT modules, resulting in a total length of approximately 10 500 m. Note that this estimated length is slightly larger than required as gaps among subpanels and modules need to be taken into consideration. The upper horizontal truss supports 50 solar subpanels. Each solar subpanel measures 200 m $\times$ 600 m and rotates around the truss via two conductive rotary joints to face the Sun accurately. The lower horizontal truss accommodates 50 antenna subpanels, each measuring 210 m $\times$ 100 m. The upper and lower south–north trusses are interconnected by 100 vertical trusses, 310 m in length. Platform service devices are installed on the solar subpanels, antenna subpanels, and the truss structures.

The specific physical configuration of the 50 solar subpanels in the proposed design is illustrated in Fig. 5. The 50 solar subpanels are arranged in the south–north direction (y-direction). Each solar subpanel includes 12 solar array modules. Each solar array module, measuring 100 m $\times$ 100 m and weighing about 3 tonnes, is manufactured and folded on Earth, launched into the orbit and then expanded to the full dimension before being assembled. The deployable truss and the film tensioning mechanism ensure that the photovoltaic film is flat after deployment. The power transmission interface handles the power output of the solar array module, and the assembly mechanism facilitates structural connections both among the solar array modules and between the solar array module and the structure truss [24]. The thin photovoltaic film is made of GaAs cells. The conversion efficiency from solar power to electric power is expected to reach 40%. With the Sun-pointing error, solar array filling factor, angle of sunlight, and space environment effects taken into consideration, the cumulative efficiency in solar power collection and conversion is estimated to be 29%, as indicated in Table I. With 1.367 kW/m$^2$ as the incident solar power density, the electric power output of each solar array module is about (1.367 kW/m$^2$) × (100 m × 100 m) × 29% = 4.0 MW. The total electric power produced by one solar subpanel is (4.0 MW per array module) × (12 array modules per subpanel) = 48 MW. The total amount of electric power produced by the entire solar panel is (48 MW per subpanel) × (50 subpanels) = 2.4 GW, approximately 29% of the 8.2 GW incident solar power. The gross weight of the solar panel is approximately 3 tonnes × 12 × 50 = 1800 tonnes.

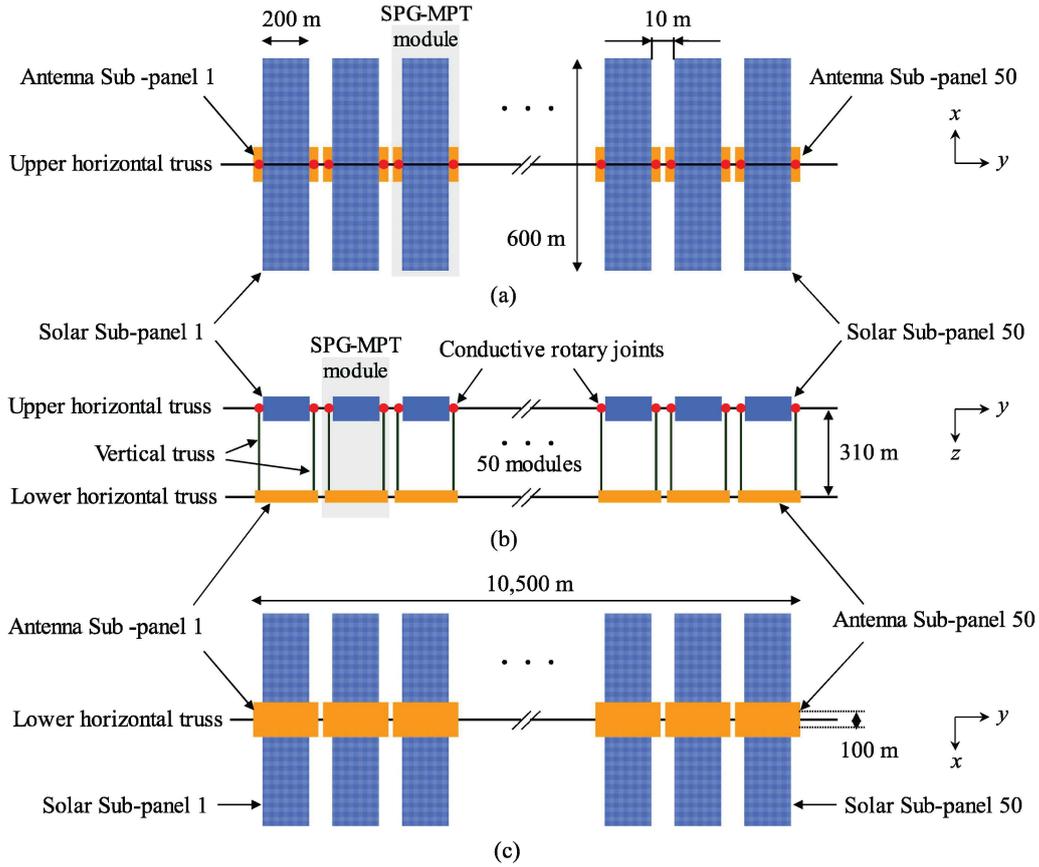

Fig. 4. Modular architecture of the solar panel and the antenna panel of the MMR-SPS concept. (a) Top view (view toward +z-direction). (b) Side view (view toward +x-direction). (c) Bottom view (view toward −z-direction).

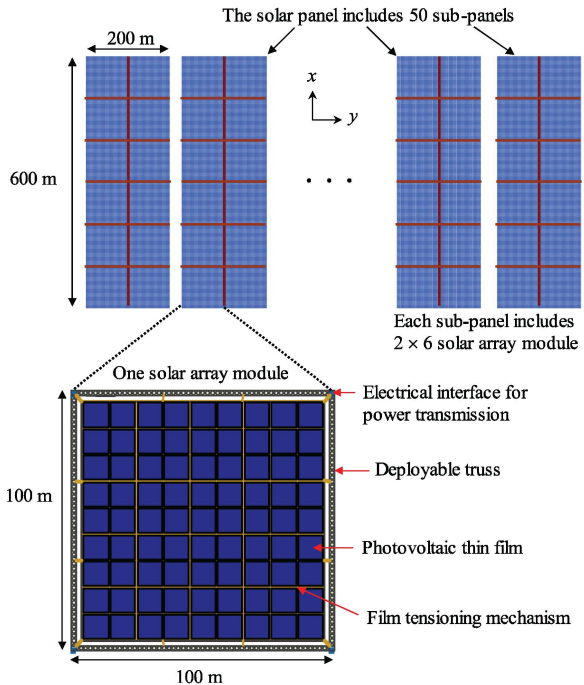

Fig. 5. Specific configuration of the solar panel.

Fig. 6 shows the physical configuration of the 50 antenna subpanels. Each antenna subpanel further includes ten array modules, with support provided by three 100 m trusses and two 210 m trusses. Each antenna array module measures

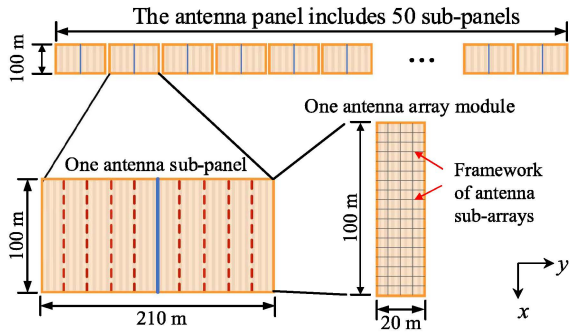

Fig. 6. Specific configuration of the antenna panel.

100 m × 20 m and weighs approximately 8.4 tonnes (with an estimated areal density of 4 kg/m$^2$). The total weight of the antenna panel is about (8.4 tonnes per array module) × (10 array modules per subpanel) × (50 subpanels) = 4200 tonnes. Like the solar array modules, the antenna array modules are manufactured on Earth, physically folded, and then launched into the orbit for assembly. The antenna panel does not have to be completely planar after deployment, as a technique called retroreflective beamforming can be employed to compensate the physical deformation of the antenna panel, which is discussed in Section III. The 48 MW dc electric power output from each solar subpanel is transmitted to the corresponding antenna subpanel with an efficiency of 89%, as estimated in Table I. Therefore, each antenna subpanel accepts (48 MW × 89%) = 42.7 MW dc power and converts it to microwave. With the dc-microwave conversion and radiation efficiency estimated to be 76%, the microwave power radiated by one antenna subpanel is approximately (42.7 MW × 76%) = 32 MW. All the radiating elements in the 50 antenna subpanels jointly construct a narrow microwave beam carrying a total power of about (32 MW × 50) = 1.6 GW toward Earth.

B. Electric PTM

Due to the SPG-MPT modular design, the electric power transmission distance from the solar panel to the antenna panel is reduced to 310 m along the vertical truss. As a result, it allows for the use of relatively shorter cables and lower voltage (5 kV), significantly simplifying the PTM system and enhancing its reliability.

The block diagram of power management in one of the 50 SPG-MPT modules is exhibited in Fig. 7. The power management scheme of the other 49 modules is identical to that in Fig. 7. The output voltages of the 12 solar array modules are 500 V. They are boosted to 5 kV before being delivered to the solar subpanel bus in order to reduce the thermal loss of the electric cables. The voltage on each solar subpanel bus is 5 kV with a current of 4800 A through each of the two rotary joints. Roller ring conductive rotary joints can be employed as they offer the benefits of low loss, high power handling capacity, high-speed stability, and extended longevity. It has been implemented on the International Space Station as the solar alpha rotary joint [26], [27]. Nevertheless, technological breakthroughs in heat dissipation, high-voltage insulation, and protection are required for roller ring conductive rotary joints to handle power levels of hundreds of kilowatts and above [12]. The 48 MW output power is transmitted to the antenna subpanel bus via the 310-m-long cables along the two vertical trusses. It is further converted back to 500 V by voltage step down converting units to supply power to the ten antenna array modules. The electric PTM is also equipped with a service subsystem shared by all modules. The overall efficiency of the PTM system, including the efficiencies of dc–dc in solar array, rotary joints and cables, dc–dc in antennas, and service devices consumption, is 89%.

C. Microwave Power Transmission

As analyzed in Section II-A, total microwave power of 1.6 GW is radiated by the antenna panel toward Earth. As illustrated in Table I, the MPT and conversion efficiencies include the atmospheric attenuation (estimated to be 95%), the microwave power collection and conversion efficiency (69%), and the dc power combination and conversion efficiency (96%) that leads to an overall dc output power on Earth to be (1.6 GW × 95% × 69% × 96%) =1 GW. One of the most important factors in these efficiencies is the beam collection efficiency (BCE), which is defined as the ratio of the microwave power collected by the receiving antenna on Earth to the power radiated by the antenna panel on the SPS (excluding the atmospheric attenuation). The BCE is expected to be higher than 90%. Given the microwave frequency to be 5.8 GHz, the dimensions of the transmitting antenna panel to be 10 500 m × 100 m, and the transmission distance to be approximately 36 000 km from GEO to Earth, a BCE of over 90% would require a receiving aperture of over 300 m × 30 000 m as estimated from the antenna theory of beamforming [28], [29]. Notice that the above high BCE requirement must be achieved by near-field focusing [30]. In fact, given the largest dimension of the solar panel $D = 10.5$ km and the wavelength $\lambda_0 = 5.17$ cm in space, the well-known far-field distance is $(2D^2/\lambda_0) = 3.9$ (million km), which is much larger than the 36 000 km transmission distance. Furthermore, the antenna array modules must be meticulously designed to mitigate their radiation along the directions of the grating lobes that arise from the intermodule gaps to ensure the achievement of 90% BCE.

Moreover, it is necessary that the normal direction of the antenna panel accurately points toward the receiving aperture on Earth (assuming broadside radiation from the antenna panel). In practice, however, deviation from the ideal status is unavoidable. Thanks to today's sophisticated satellite control technology [31], it is highly possible that the attitude deviation of solar panel and antenna panel does not exceed 0.1°. Nevertheless, an attitude deviation in the range of [−0.1°, 0.1°] still significantly influences the BCE. An angular error of 0.1° (or 0.0017 radian) transforms to a pointing error of (36 000 km) × (0.0017 radian) = 62.8 km on Earth, which is way beyond the size of the receiving aperture. Therefore, the antenna panel needs to employ phased arrays to steer the power beam with much higher accuracy.

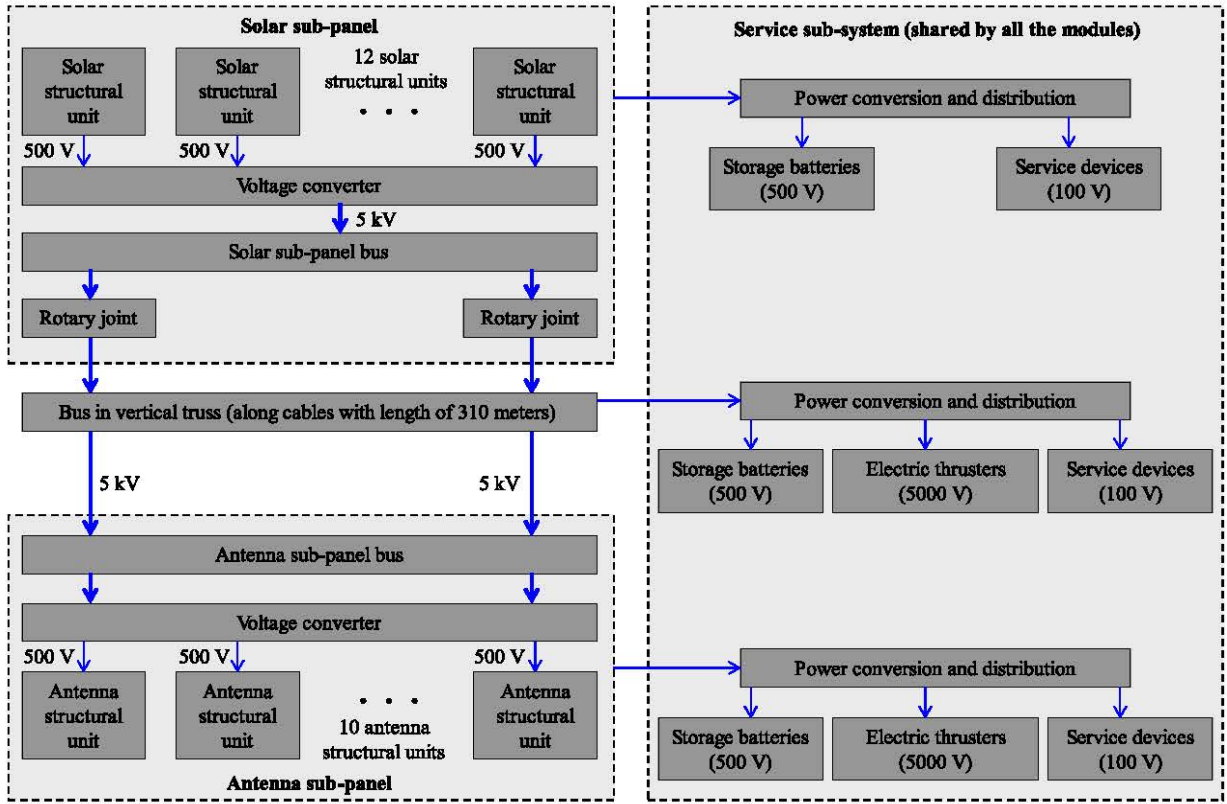

Fig. 7. Block diagram of power management in one of the 50 modules.

In a conventional SPS concept, such as the MR-SPS, the typical beam steering accuracy requirement is generally considered to be 0.0005° so that the deviation of the footprint of the microwave power beam from GEO can be constrained within 300 m on Earth. This is expected to be achieved by the transmitting antenna in the form of a circular phased array with a diameter of about 1 km. However, in the proposed MMR-SPS concept, the much larger size of the transmitting antenna on the long side (10.5 km along the $y$-direction, as shown in Fig. 4) results in a concentrated power distribution within a much narrower region along this direction. This demands a significantly higher beam steering accuracy along this specific direction. However, on a positive note, the extended 10.5-km-long side of the transmitting antenna allows for the incorporation of more phase-controlled antenna elements along this direction. This capability enables the implementation of a much more precise beam-control mechanism, enhancing the overall beam steering accuracy along the direction.

Referring to Miyakawa et al. [32], the 1-D beam pointing error of a phased array antenna with beam direction control using monopulse method can be calculated by (assuming all radiating elements have the same excitation amplitude)

$$\Delta\theta = \frac{2\lambda_0}{(\pi d \cos\theta_0) M^{1.5}} \Phi \qquad (1)$$

where $\Delta\theta$ is the root-mean-square (rms) beam pointing error, $M$ is the total number of linear radiating elements, $\Phi$ is the rms error of the feeding phases of the transmitting antenna, $d$ is the spacing of the radiating elements, $\lambda_0$ is the wavelength, and $\theta_0$ is the beam pointing direction from the normal direction of the power-transmitting antenna. For the MMR-SPS concept, if the regular half-wavelength element spacing is adopted (i.e., if $d = 2.59$ cm), a total of $M = 386\,000$ radiating elements are distributed along the long side. Let $\cos\theta_0 = 1$ and $\Phi = 20°$ (rms), the ideal beam pointing accuracy $\Delta\theta$ of the transmitting antenna along the long side direction is approximately $1.1 \times 10^{-7}$° (rms). However, for the half-wavelength spacing of radiating elements, there is a linear distribution of 3860 elements along the short (100 m) side of the antenna panel, leading to a total of $(386\,000 \times 3860) \approx 1.5$ (billion) radiating elements. Applying independent phase control of 1.5 billion antenna elements leads to a formidably high cost. Therefore, it is practical to divide the entire antenna array into multiple subarrays, apply identical feed phase to all the elements within each subarray, and apply independent phase control to the subarrays. Specifically, with the size of the subarray to be 5 m × 5 m, the total number of independent phase control circuits is reduced to (2000 × 20) = 40 000. In this case, $d = 5$ m and a total of $M = 2000$ antenna subarrays are distributed along the long side. With $\cos\hat{\theta} = 1$ and $\Phi = 20°$ (rms), the beam pointing accuracy $\Delta\theta$ given by (1) is approximately $1.5 \times 10^{-6}°$ rms, i.e., less than 10 m deviation for the 36 000 km transmission distance.

## III. RETROREFLECTIVE BEAMFORMING OF MPT

The beam pointing accuracy estimation given by (1) only considers the random feeding phase error. Many other practical factors might also significantly affect the beam steering accuracy and eventually affect the BCE. One of the most important factors is the antenna position deviation.

While the SPS is positioned in the GEO, achieving and maintaining a completely flat and stationary kilometer-large antenna array at all times is impractical. Assembling errors can introduce deviations in the attitude and relative positions of the antenna modules, impacting the beamforming quality and pointing accuracy of the MPT system. In addition, various environmental factors also contribute to antenna position deviations and aperture deformation. For example, extreme temperature fluctuations cause expansion and contraction of the antenna structure. External forces, such as solar radiation pressure, result in distortion and bending. Potential collisions with small meteoroids or micrometeorites lead to surface deformations and vibrations. The combined effect of these factors results in the misalignment of the antenna array aperture from its ideal planar state, detrimentally affecting beamforming performance and pointing accuracy. How to compensate for these adverse effects stands as a fundamental challenge in the MPT design for SPS. One way of achieving the compensation is by detecting the antenna position deviation with sensors and then making shape calibrations, which obviously increases the system complexity significantly. A framework to reconstruct the shape of flexible arrays via mutual coupling measurements while requiring no additional sensors is proposed in [33], which provides a promising approach for compensating the antenna position deviations effectively. This article focuses on another technique to address this challenge, i.e., the retroreflective beamforming [30], [34], [35].

In a retroreflective beamforming scheme, MPT is accomplished via the following two steps.

*Step 1:* The microwave power receiver broadcasts a pilot signal as the "request for wireless power." The phase distribution of the pilot signal arriving at the antenna array of the microwave power transmitter is detected.

*Step 2:* By feeding the power-transmitting array with a phase distribution conjugate to that of the received pilot signal, the microwave power transmitter generates a microwave power beam toward the power receiver.

To simulate the power density distribution over the receiving aperture and estimate the effect of the antenna position deviation on the BCE, a calculation model, as illustrated in Fig. 8, is used. In the model, the SPS includes a planar antenna array with $M \times N$ phased antenna elements in the $x$–$y$ plane, with $M$ rows and $N$ columns. $\mathbf{r}'_{mn}$ is the position vector of the phase center of the antenna element in the $m$th row and the $n$th column. $\mathbf{r}$ is the position vector of an observation point on the power-receiving aperture. The phased transmitting array generates a microwave power beam focused on the location from which the pilot signal is broadcasted. This location is represented by the position vector $\mathbf{r}_0$, which is usually the center of the receiving aperture.

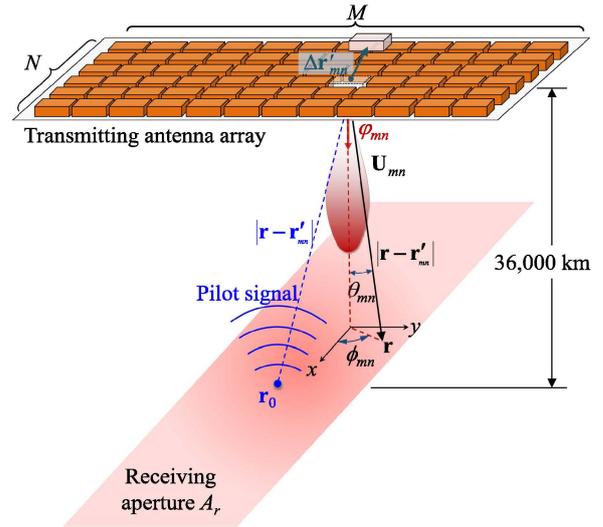

Fig. 8. Retroreflective beamforming model of MPT in MMR-SPS (not to scale).

It is important to notice that $\mathbf{r}$ (as well as $\mathbf{r}_0$) resides in the near-field region of the planar antenna array of the SPS. However, for each single antenna element (an antenna subarray with dimensions 5 m × 5 m) in the antenna panel, $\mathbf{r}$ is in the far-field region. Therefore, the total electric field at an observation point $\mathbf{r}$ can be written as

$$\mathbf{E}(\mathbf{r}) = \sum_{m=1}^{M}\sum_{n=1}^{N} \frac{A_{mn} e^{j\varphi_{mn}} e^{-jk_0|\mathbf{r}-\mathbf{r}'_{mn}|}}{|\mathbf{r}-\mathbf{r}'_{mn}|} \mathbf{U}_{mn}(\theta_{mn}, \phi_{mn}) \quad (2)$$

where $A_{mn}$ and $\varphi_{mn}$ represent the magnitude and phase of the excitation of the antenna element with indices $(m, n)$. $k_0 = 2\pi/\lambda_0$ is the wavenumber in free space. $\mathbf{U}_{mn}$ is a polarization vector representing the normalized field contribution of the antenna element with indices $(m, n)$ to the total $\mathbf{E}$ field. $\mathbf{U}_{mn}$ depends on the direction of the observation point $\mathbf{r}$ relative to the phase center $\mathbf{r}'_{mn}$, which is a function of the angles $\theta_{mn}$ and $\phi_{mn}$, as illustrated in Fig. 8. Although the field contribution from each antenna element is calculated with far-field approximation, (2) is different from the commonly used far-field calculation model in the following aspects.

1) The phase delay $k_0|\mathbf{r} - \mathbf{r}'_{mn}|$ in the exponential propagation factor is evaluated rigorously, instead of assuming a linear phase variation over the antenna array aperture in the far-field approximation.

distribution is disturbed as

$$\mathbf{E}_{\text{deform}}(\mathbf{r}) = \sum_{m=1}^{M}\sum_{n=1}^{N} \frac{A_{mn} e^{-jk_0(|\mathbf{r}-\mathbf{r}'_{mn}-\Delta\mathbf{r}'_{mn}|-|\mathbf{r}_0-\mathbf{r}'_{mn}|)}}{|\mathbf{r}-\mathbf{r}'_{mn}-\Delta\mathbf{r}'_{mn}|} \mathbf{U}_{mn}$$

(5)

where $\Delta\mathbf{r}'_{mn}$ represents the position deviation of the antenna element with indices $mn$, as illustrated in Fig. 8. To quantify the beamforming performance degradation, a degradation factor can be introduced as follows:

$$\text{DF}_{\text{deform}} = \frac{\iint_{A_r} |\mathbf{E}_{\text{deform}}(\mathbf{r})|^2 dS}{\iint_{A_r} |\mathbf{E}_{\text{ideal}}(\mathbf{r})|^2 dS} \qquad (6)$$

which can be considered as a decrease factor in BCE.

To compensate for the effect of deformation, retroreflective beamforming is employed, which fixes the excitation phase as

$$\varphi_{mn} = k_0 \left|\mathbf{r}_0 - \mathbf{r}'_{mn} - \Delta\mathbf{r}'_{mn}\right|. \qquad (7)$$

The **E**-field distribution after compensation of the retroreflective beamforming becomes

$$\mathbf{E}_{\text{retro}}(\mathbf{r}) = \sum_{m=1}^{M}\sum_{n=1}^{N} \frac{A_{mn} e^{-jk_0(|\mathbf{r}-\mathbf{r}'_{mn}-\Delta\mathbf{r}'_{mn}|-|\mathbf{r}_0-\mathbf{r}'_{mn}-\Delta\mathbf{r}'_{mn}|)}}{|\mathbf{r}-\mathbf{r}'_{mn}-\Delta\mathbf{r}'_{mn}|} \mathbf{U}_{mn}$$

(8)

The degradation factor after compensation, denoted by $\text{DF}_{\text{retro}}$, can be calculated as

$$\text{DF}_{\text{retro}} = \frac{\iint_{A_r} |\mathbf{E}_{\text{retro}}(\mathbf{r})|^2 dS}{\iint_{A_r} |\mathbf{E}_{\text{ideal}}(\mathbf{r})|^2 dS} \qquad (9)$$

The antenna position deviations can be divided into two types: random position errors and overall aperture deformations. All these deviations are most likely caused by shifts of the antenna modules in the $z$-direction [36], [37].

For random positional errors, two cases are considered here: uniform distributions and normal distributions of the position error in the $z$-directions. In other words

$$\Delta\mathbf{r}'_{mn} = \Delta z_{mn} \hat{\mathbf{z}} \qquad (10)$$

where $\Delta z_{mn}$ is a random variable with uniform or normal distributions.

For overall aperture deformations, the bent and bow deformation models proposed in [36] and the sine and saddle models proposed in [38] are adopted in the simulation, as shown in Fig. 10.

Fig. 11 shows the degradation factors due to the two types of random position errors and the bent aperture deformations along the $x$ and $y$ directions as functions of the *position deviation amount*. For random positional errors, the position deviation amount is defined as the standard deviation of the random variable $\Delta z_{mn}$. For the overall aperture deformations, the position deviation amount is defined as the difference between the maximum and minimum of $\Delta z_{mn}$.

From the results in Fig. 11, it is seen that without compensation of the retroreflective beamforming, a bent deformation as small as 4 cm along the 10-km-long side ($y$-direction) causes a degradation factor $\text{DF}_{\text{deform}} = 91\%$.

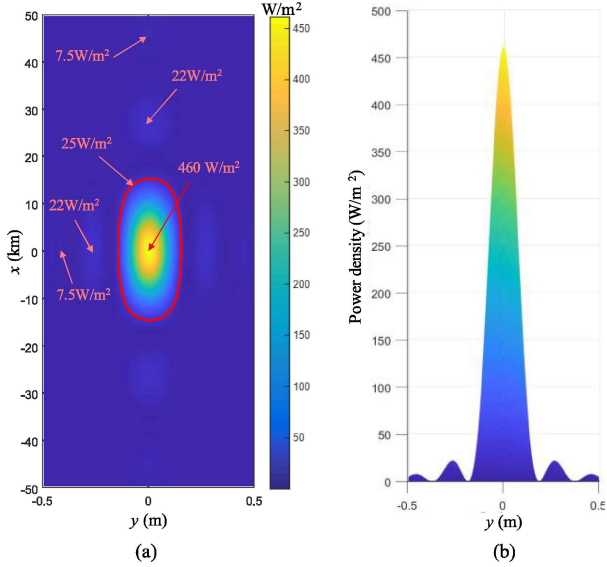

Fig. 9. Power density distribution in the idea case (not to scale). (a) Over the receiving aperture on the ground. (b) Across the aperture with fixed $x = 0$.

2) The distances $|\mathbf{r} - \mathbf{r}'_{mn}|$ in the denominator is evaluated rigorously, instead of assuming the same distance for all $(m, n)$ in the far-field approximation.
3) $\mathbf{U}_{mn}$ is different for different indices $(m, n)$ with the angles $\theta_{mn}$ and $\phi_{mn}$ evaluated rigorously, instead of assuming the same direction of $\mathbf{r}$ relative to $\mathbf{r}'_{mn}$ for all $(m, n)$ in the far-field approximation.

With the above improvements in the calculation model, the near-field effect is taken into account in (2).

In the ideal case, the excitation phase $\varphi_{mn}$ is determined with the retroreflective beamforming scheme based on the principle of phase conjugation, which gives

$$\varphi_{mn} = k_0 \left|\mathbf{r}_0 - \mathbf{r}'_{mn}\right| \qquad (3)$$

so that the microwave power beams from all the antenna elements superpose in phase on the desired observation point $\mathbf{r}_0$. In this case, the **E**-field distribution over the receiving aperture can be written as

$$\mathbf{E}_{\text{ideal}}(\mathbf{r}) = \sum_{m=1}^{M}\sum_{n=1}^{N} \frac{A_{mn} e^{-jk_0(|\mathbf{r}-\mathbf{r}'_{mn}|-|\mathbf{r}_0-\mathbf{r}'_{mn}|)}}{|\mathbf{r}-\mathbf{r}'_{mn}|} \mathbf{U}_{mn} \quad (4)$$

With the size of phase-controlled antenna elements to be 5 m × 5 m, $M = 2000$, and $N = 20$, the simulated power density distribution on the ground (see Fig. 9) reveals a main lobe zone of about 30 km × 300 m with a BCE of 85% and the highest power density reaching about 460 W/m². The first sidelobe zone and second sidelobe zone have the highest power densities of approximately 22 W/m² and 7.5 W/m²;, respectively. To achieve a higher BCE, the rectenna array on the ground is designed to cover an area of 60 km × 800 m.

However, in the presence of transmitting aperture deformation and element position deviations, there will be a degradation of the beamforming performance as the **E**-field

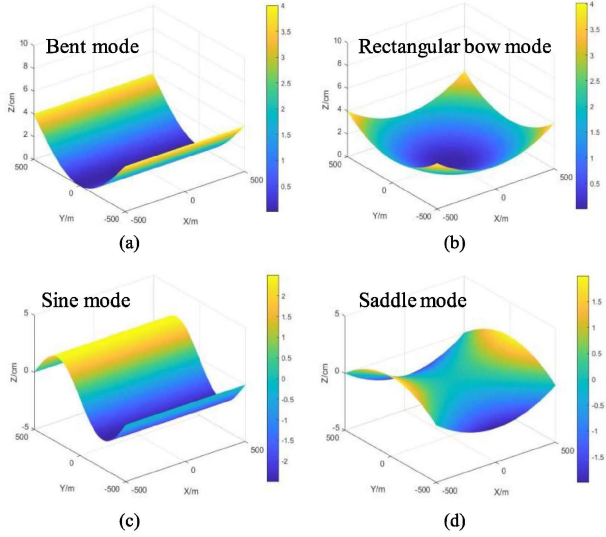

Fig. 10. Four deformation models of large-scale arrays.

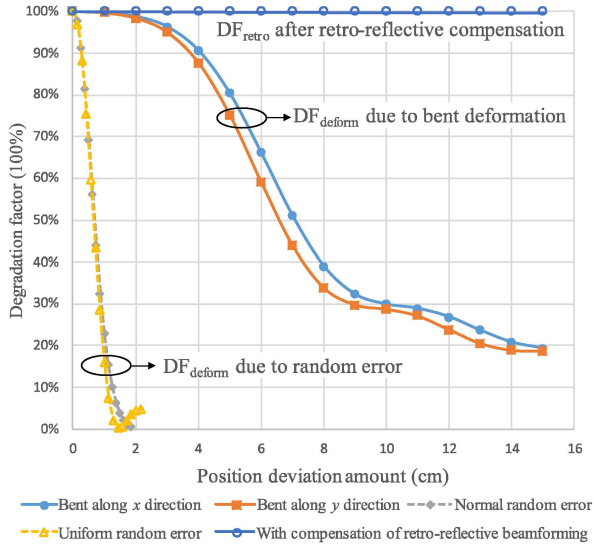

Fig. 11. Degradation factor of the 10 km × 100 m transmitting antenna array of MMR-SPS due to different types of the antenna position deviations.

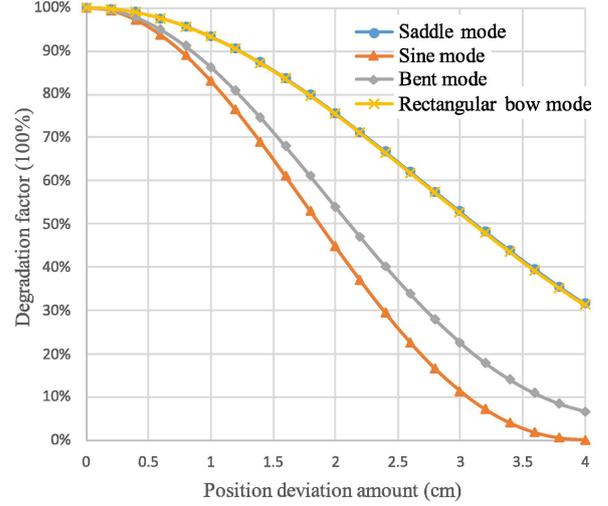

Fig. 12. Degradation factor of a 5 m × 5 m antenna element due to different types of the aperture deformations.

The same amount of bent deformation along the 100-m-long side leads to $DF_{deform} = 88\%$. When the bent deformation amount exceeds 7 cm, $DF_{deform}$ below 50% is observed, indicating a substantial reduction of more than 50% in BCE. Moreover, for random antenna position errors, a tolerance of only 3 mm deviation is permissible if a degradation factor of 90% or above is to be maintained. Nevertheless, with the retroreflective beamforming compensations, the degradation factor $DF_{retro}$ is maintained at almost 100% for all deviation amounts. Therefore, for the exceptionally large-scale antenna array of SPS, retroreflective beamforming plays a crucial role in achieving highly efficient MPT.

It is important to note that the retroreflective beamforming scheme cannot compensate for degradation of the beamforming performance caused by the deformation within a single-fed antenna element (subarray). Given the antenna element size of 5 m × 5 m, it is challenging to completely avoid the inner deformation. Fig. 12 illustrates the simulated degradation factor due to four types of aperture deformations of the antenna elements. It is seen that, to maintain a degradation factor of 90% or above, the overall aperture deviation of the 5 m × 5 m antenna elements should not exceed 8 mm.

While reducing the antenna element size makes it easier to minimize the inner deformation, it comes with the price of requiring more antenna elements and circuits. This, in turn, significantly increases the system's complexity and cost.

In the above analysis, it is assumed that the pilot signal has the same frequency as the microwave power beam. In this case, a time-division duplex scheme must be used to provide isolation of the transmitted power from the received pilot signal. In practice, frequency-division duplexing can also be used. Fig. 13 illustrates a configuration scheme of the retroreflective beamforming circuits with frequency-division duplexing. The frequency of the pilot signal is 2.9 GHz, half of the frequency of the microwave power beam. Each antenna element contains a 5 m × 5 m power-transmitting subarray and a small pilot signal-receiving antenna embedded within the subarray. For accurate retroreflective beamforming control, the phase center of the power-transmitting subarray should overlap with the phase center of the pilot signal-receiving antenna.

The received pilot signal is filtered and amplified, and then downconverted to baseband for digital processing. The digital processing is performed by a phase conjugation circuit, which converts the baseband pilot signal with phase $\varphi'_{mn}$ to a conjugate baseband output with phase $\varphi_{mn} = -2\varphi'_{mn}$. The conjugate baseband signal is then upconverted to 5.8 GHz, amplified and fed to the power-transmitting subarray. The power-transmitting circuit, pilot signal-receiving circuit, and the phase conjugation circuit are installed on the back face of each antenna element (subarray), all synchronized by a local oscillator (LO) signal for precise timing.

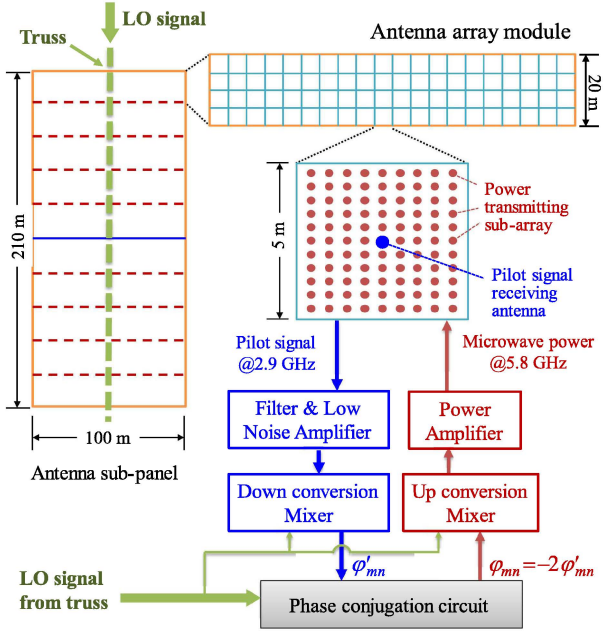

Fig. 13. Configuration scheme of the retroreflective beamforming antennas and circuits.

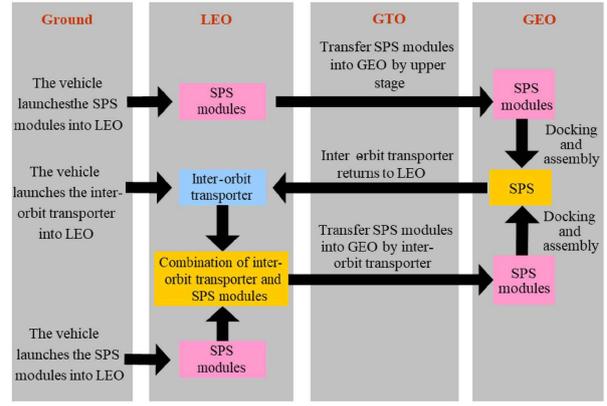

Fig. 14. Schematic diagram of GEO assembly and transportation mode.

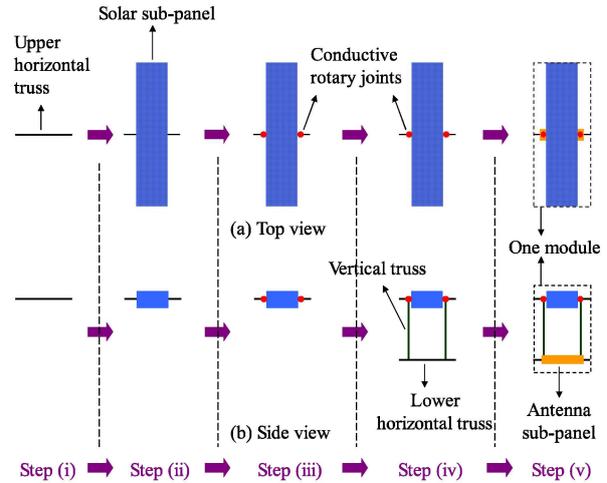

Fig. 15. Assembly sequence of an SPG-MPT module of the MMR-SPS. (a) Top view. (b) Side view.

The LO signal originates from a single sinusoidal signal generator and is distributed to each antenna array module via the lower south–north truss, as shown in Fig. 13. As one of the major technical challenges of the SPS design, distribution of the LO signal for the retroreflective beamforming circuits calls for a phase-stabilized frequency transfer network spanning kilometers, which is still undergoing investigation.

## IV. TRANSPORTATION AND ASSEMBLY IN ORBIT

The obviously enormous mass and volume of SPS pose significant challenges for its transportation and assembly in orbit, necessitating division into multiple modules for launch into orbit based on the transport capacity of the launch vehicle. MMR-SPS adopts the GEO assembly and transportation mode. The process involves utilizing heavy-lift rockets to launch the SPS modules into low Earth orbit (LEO). Then, the upper stage of the rocket or an interorbit transporter is employed to transport all the SPS modules to GEO, where assembly takes place (refer to Fig. 14).

Key features of this approach include conducting all assembly processes in GEO, which significantly reduces the risk of collisions with space debris. This is due to the substantially lower concentration of debris present in GEO compared with lower orbits, thereby enhancing the safety and reliability of the assembly operations. The modular transportation strategy results in a relatively low transportation capacity requirement for the interorbit transporter, albeit necessitating more interorbit transporters for support. The impact of Earth's radiation belt on the SPS modules during orbit transfer is reduced by the protective shield of the interorbit transporter. Moreover, the interorbit transporter can achieve long-term reusability through propellant refueling and the solar array replacement in LEO.

The MMR-SPS comprises solar array modules, truss modules, antenna modules, and other modules designed for convenient assembly in space. The pivotal aspect of this updated concept is that it consists of 50 identical SPG-MPT modules, and all modules are mutually independent.

The assembly sequence of an individual module is depicted in Fig. 15. First, the solar array modules are deployed and assembled in orbit, followed by the sequential assembly of the corresponding conductive rotary joints and upper south–north trusses. Second, two connecting perpendicular trusses are assembled to the south–north truss and deployed in the z-direction. Third, the deployed lower south–north truss is assembled between the two connecting trusses. Fourth, the antenna modules are deployed and assembled based on the lower south–north truss. Finally, all cables and electrical equipment are assembled.

Space assembly robots are employed to complete the assembly, providing the functions of grabbing assembly modules, transporting modules to specific locations, assembling modules, and assisting in the deployment of modules. The space assembly robot system mainly includes platform mobile robots and free maneuvering robots. Platform mobile robots move and complete space assembly and maintenance operations through a moving platform on a truss or other structures. The mechanical arm operation system of the International Space Station is a typical platform mobile robot [39]. Almost all large structures of the space station are assembled through the mechanical arm operation system. Free maneuvering robots have the ability to move unhindered along orbits. They are equipped with a mechanical arm system that can achieve rapid transportation and assembly of payloads, facilitating the efficient construction of ultralarge space structures. Currently, research in free maneuvering robots is still in its early stages. All the robots are launched into the orbit together with the assembly modules. After that, they operate independently. Considering the assembly efficiency, it is estimated that a minimum of several dozen robots must operate simultaneously.

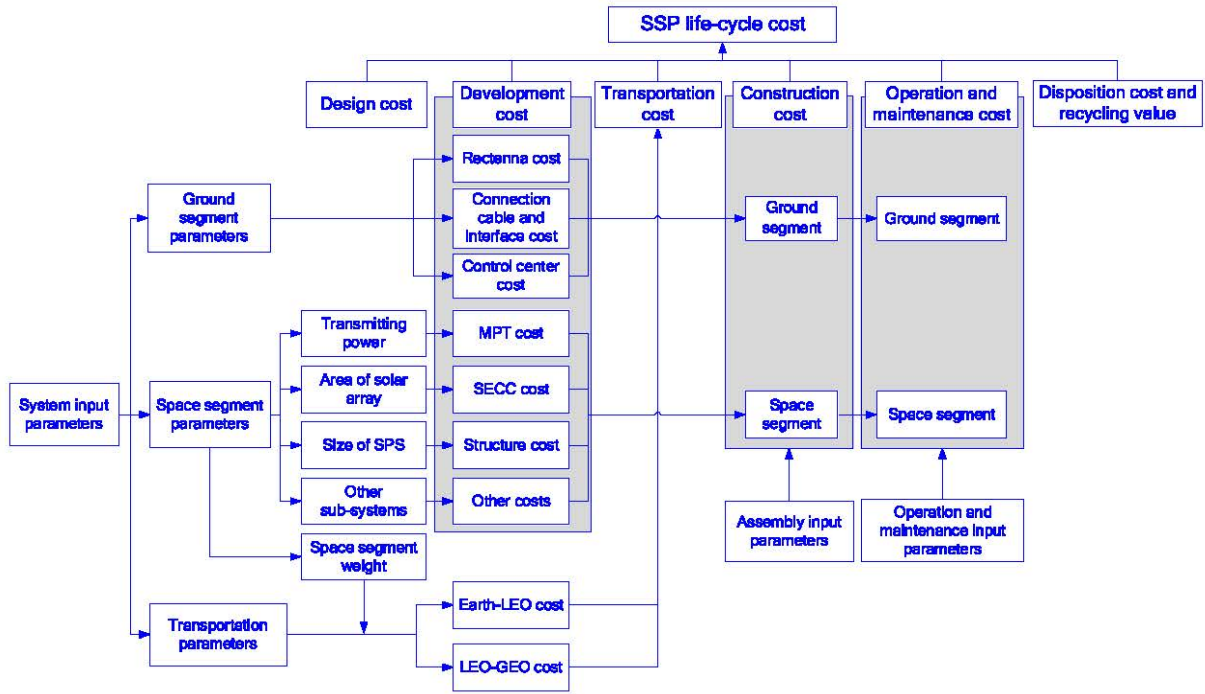

Fig. 16. Flowchart of life-cycle cost analysis of an SSP.

Since one solar subpanel corresponds to one antenna subpanel, the power generated by the solar subpanel can be transmitted to the antenna subpanel through two cables installed on the perpendicular trusses. This approach offers significant advantages for on-orbit assembly. Once an individual SPG-MPT module is assembled, it can operate and transmit power immediately. Subsequently, all SPG-MPT modules are repeatedly assembled, and the construction of the entire SPS system involves the continuous addition of modules in the $y$-direction. Accidental damage, such as debris impact, will not affect the operation of the entire system, and each module can be maintained and repaired independently.

## V. LIFE-CYCLE COST ANALYSIS

The life-cycle cost analysis of an SSP system is conducted following the flowchart in Fig. 16. The entire SSP system comprises a space segment and a ground segment. The analysis process begins by inputting the basic system parameters and selecting and analyzing the relevant parameters for both the ground and space segments. Subsequently, the design cost and development cost are calculated, followed by the computation of transportation costs based on specified parameters. The construction cost is determined according to assembly parameters, and operation and maintenance costs are calculated based on the related parameters. Estimations are made for disposition costs and recycling value. Ultimately, the sum of these costs represents the life-cycle cost of an SSP system.

A cost analysis of the original MR-SPS concept using the above approach is provided in [24]. An update of the cost data for the proposed MMR-SPS is presented in Tables II and III, showing the primary cost analysis results for the space segment and the ground segment, respectively. The calculated total cost of an MMR-SPS over a 30-year period is approximately 17.3 billion U.S. dollars. Assuming a continuous output power of 1 GW, the total life-cycle energy supplied by an SPS is about 262.8 billion kWh. The cost per kWh is calculated to be approximately 6.6 U.S. cents. Compared with the total cost estimation in [24], the cost estimation for MMR-SPS is lowered by more than 30%. The cost data also serve as the design requirement for future SPS development efforts and must be continuously amended as research and technology progress.

## VI. CONCLUSION

In summary, this article introduces and analyzes the primary scheme of an MMR-SPS concept, outlining the key parameters of a typical 1 GW MMR-SPS, as shown a  SECC — solar energy collection and conversion.   b  PTM — power transmission and management.
c  MPT — microwave power transmission.   d  AOC — structure, attitude, and orbit control.
e  TM — thermal management.   f  ISRM — information and system operation management.

TABLE II
Cost Analysis Result of the Space Segment (Million U.S. Dollars)

| | Design | Development | Transportation ($1000/kg) | Construction | Operation and maintenance | Disposition and recycling | Total |
|---|---|---|---|---|---|---|---|
| SECC[a] subsystem | 7 | 2570 | 850 | 855 | 226 | 255 | 4763 |
| PTM[b] subsystem | 14 | 585 | 600 | 355 | 255 | 193 | 2002 |
| MPT[c] subsystem | 7 | 1800 | 2000 | 783 | 212 | 600 | 5402 |
| Structure subsystem | 14 | 410 | 770 | 712 | 42 | −975 | 973 |
| AOC[d] subsystem | 14 | 455 | 40 | 70 | 770 | 8 | 1357 |
| TM[e] subsystem | 14 | 56 | 105 | 70 | 85 | 28 | 358 |
| ISRM[f] subsystem | 21 | 140 | 52 | 35 | 85 | 14 | 347 |
| Total | 91 | 6016 | 4417 | 2880 | 1675 | 123 | 15 202 |

TABLE III
Cost Analysis Result of the Ground Segment (Million U.S. Dollars)

| | Design | Development | Construction | Operation and Maintenance | Disposition and Recycling | Total |
|---|---|---|---|---|---|---|
| Rectenna | 7 | 1420 | 282 | 300 | −28 | 1981 |
| Connect Cable and Interface | 1.4 | 14 | 7 | 21 | −2.8 | 40.6 |
| Control Center | 1.4 | 5.7 | 1.4 | 42 | −1.4 | 49.1 |
| Total | 9.8 | 1439.7 | 290.4 | 363 | −32.2 | 2070.7 |

[a]SECC–solar energy collection and conversion.
[b]PTM–power transmission and management.
[c]MPT–microwave power transmission.
[d]AOC–structure, attitude, and orbit control.
[e]TM–thermal management.

TABLE IV
Summary of a 1 GW MMR-SPS

| System | Parameter | Value |
|---|---|---|
| SPS system | Orbit | GEO |
| | Supply power /GW | ~1 |
| | Total efficiency | ~12.3% |
| | Total mass /t | ~9160 |
| | Life-cycle cost/M$ | ~17300 |
| Solar Energy Collection and Conversion subsystem | Efficiency of PV cell | ~40% |
| | Number of solar sub-panels | 50 |
| | Number of solar array modules | 600 |
| | Area of solar panel /km² | ~6 |
| | Output power /GW | ~2.4 |
| | Voltage of solar array module/V | ~500 |
| | Mass/t | ~1800 |
| MPT subsystem | Frequency/GHz | 5.8 |
| | Efficiency of WPT | ~54% |
| | Size of transmission antenna/km | 0.1×10.5 |
| | Number of antenna subpanels | 50 |
| | Size of antenna array modules/m | 5 |
| | Deform of antenna array module/mm | < 8 |
| | Transmission power of an antenna subpanel/MW | 32 |
| | Mass/t | ~4200 |
| PTM subsystem | Voltage of main bus/kV | 5 |
| | Number of rotary joints | 100 |
| | Mass/t | ~1360 |
| Other subsystem | Mass/t | 1800 |

in Table IV. The innovative distribution layouts of solar subpanels and antenna subpanels offer a solution to the challenges associated with long-distance, high-voltage electric power transmission in space. Each antenna subpanel, paired with the corresponding solar subpanel, constitutes an independent SPG-MPT module, leading to a modular SPS configuration. This modularity effectively reduces the complexity and weight of the space PTM subsystem.

A distinctive feature of the proposed MMR-SPS concept is the elongated aperture of the transmitting antenna panel, posing challenges in high-precision MPT. To address this, the retroreflective beamforming scheme is introduced. With a numerical calculation model, retroreflective beamforming proves to be necessary and effective in compensating for the impact of antenna position deviation and aperture deformation.

The assembly of the entire SPS is also facilitated by the modular configuration, allowing each module to operate independently. Moreover, the modular nature enables each unit to transmit power immediately after assembly, improving the utilization efficiency of the entire SPS.

REFERENCES

[1] P. E. Glaser, "Power from the Sun: Its future," Science, vol. 162, pp. 867–886, 1968.
[2] FNC/LE, Frazer Nash Consultancy and London Economics, "Space-based solar power: A future source of energy for Europe," 2022. [Online]. Available: https://astrostrom.ch/docs/Brochure(EN)_Space-Based_Solar_Power_A_Future_Source_of_Energy_for_Europe.pdf
[3] Roland Berger, "Space-based solar power, can it help to decarbonize Europe and make it more energy resilient," Munich, Germany, 2022.
[4] A. Fikes et al., "The Caltech space solar power demonstration one mission," in Proc. IEEE Int. Conf. Wireless Space Extreme Environ., Winnipeg, MB, Canada, 2022, pp. 18–22.

[5] S. Mihara et al., "The road map toward the SSPS realization and application of its technology," in *Proc. 69th Int. Astronaut. Congr.*, Bremen, Germany, 2018.
[6] M. Soltau et al., "The U.K. space energy initiative—Towards a practical space based power system for the net zero era," in *Proc. 72th Int. Astronaut. Congr.*, Paris, France, 2022.
[7] J. M. Choi, "Conceptual design of Korean space solar power satellite," in *Proc. 70th Int. Astronaut. Congr.*, Washington DC, USA, 2019.
[8] J. C. Mankins, "IAA decadal assessment of space solar power: A progress report," in *Proc. 71th Int. Astronaut. Congr.*, Dubai, UAE, 2021.
[9] B. Strassner and K. Chang, "Microwave power transmission: Historical milestones and system components," *Proc. IEEE*, vol. 101, no. 6, pp. 1379–1396, Jun. 2013.
[10] "Guidance on frequency ranges for operation of wireless power transmission via radio frequency beam for mobile/portable devices and sensor networks," vol. 9, 2022. [Online]. Available: https://www.itu.int/dms_pubrec/itu-r/rec/sm/R-REC-SM.2151-0-202209-I!!PDF-E.pdf
[11] E. Gholdston, J. Hartung, and J. Friefeld, "Current status, architecture, and future technologies for the International Space Station electric power system," *IEEE Aerosp. Electron. Syst. Mag.*, vol. 11, no. 2, pp. 25–30, Feb. 1996.
[12] X. B. Hou, L. Wang, and Q. Li, "Review of key technologies for high-voltage and high-power transmission in space solar power station," *Trans. China Electrotech. Soc.*, vol. 33, no. 14, pp. 3385–3395, 2018.
[13] X. B. Hou and L. Wang, "Study on multi-rotary joints space power satellite concept," *Aerosp. China*, vol. 19, pp. 19–26, 2018.
[14] J. C. Mankins, "SPS-ALPHA Mark-III and an achievable roadmap to space solar power," in *Proc. 71th Int. Astronaut. Congr.*, Dubai, UAE, 2021.
[15] Y. Yang, Y. Zhang, B. Duan, D. Wang, and X. Li, "A novel design project for space solar power station (SSPS-OMEGA)," *Acta Astronautica*, vol. 121, pp. 51–58, 2016.
[16] I. Cash, "CASSIOPeiA—A new paradigm for space solar power," *Acta Astronautica*, vol. 159, pp. 170–178, 2019.
[17] S. Sasaki et al., "A new concept of solar power satellite: Tethered-SPS," *Acta Astronautica*, vol. 60, no. 3, pp. 153–165, Feb. 2007.
[18] M. Arya, N. Lee, and S. Pellegrino, "Ultralight structures for space solar power satellites," in *Proc. 3rd AIAA Spacecraft Struct. Conf.*, Jan. 2016.
[19] M. A. Marshall, A. Goel, and S. Pellegrino, "Power-optimal guidance for planar space solar power satellites," *J. Guid., Control, Dyn.*, vol. 43, no. 3, pp. 518–535, 2020.
[20] X. Hou et al., "MMR-SPS: An updated concept design on MR-SPS," in *Proc. 74th Int. Astronautical Congr.*, Baku, Azerbaijan, 2023.
[21] "Best research-cell efficiency chart," Accessed on: Jun. 1, 2024. [Online]. Available: https://www.nrel.gov/pv/cell-efficiency.html
[22] C. T. Rodenbeck et al., "Microwave and millimeter wave power beaming," *IEEE J. Microw.*, vol. 1, no. 1, pp. 229–259, Jan. 2021.
[23] E. Rodgers et al., "Space based solar power," 2024, Accessed on: May 17, 2024. [Online]. Available: https://ntrs.nasa.gov/citations/20230018600
[24] X. Hou, L. Wang, and X. Zhang, *An Introduction to Space Solar Power Stations*. Beijing, China: China Astronautics Publishing House, 2020.
[25] A. Bermudez-Garcia, P. Voarino, and O. Raccurt, "Environments, needs and opportunities for future space photovoltaic power generation: A review," *Appl. Energy*, vol. 290, May 2021, Art. no. 116757, doi: 10.1016/j.apenergy.2021.116757.
[26] E. P. Harik, J. McFatter, D. J. Sweeney, C. F. Enriquez, D. M. Taylor, and D. S. McCann, "The international space station solar alpha rotary joint anomaly investigation," May 2010, Accessed on: Jun. 5, 2024. [Online]. Available: https://ntrs.nasa.gov/citations/20100021920
[27] S. Loewenthal, C. Allmon, C. Reznik, J. Mcfatter, and R. E. Davis, "Space station solar array joint repair," *Mater. Perform. Characterization*, vol. 4, no. 1, pp. 200–208, Nov. 2015.
[28] W. C. Brown and E. E. Eves, "Beamed microwave power transmission and its application to space," *IEEE Trans. Microw. Theory Techn.*, vol. 40, no. 6, pp. 1239–1250, Jun. 1992.
[29] C. A. Balanis, *Antenna Theory: Analysis and Design*, 3rd ed. New York, NY, USA: Wiley, 2005.
[30] X. Wang, X. Hou, L. Wang, and M. Lu, "Employing phase-conjugation antenna array to beam microwave power from satellite to Earth," in *Proc. IEEE Int. Conf. Wireless Space Extreme Environ.*, Orlando, FL, USA, 2015, pp. 1–5.
[31] B. Wu, D. Wang, and E. K. Poh, "High precision satellite attitude tracking control via iterative learning control," *J. Guid., Control, Dyn.*, vol. 38, no. 3, pp. 528–533, Mar. 2015.
[32] T. Miyakawa et al., "Preliminary experimental results of beam steering control subsystem for solar power satellite," in *Proc. 63rd Int. Astronaut. Congr.*, Naples, Italy, 2012.
[33] A. Fikes, O. S. Mizrahi, and A. Hajimiri, "A framework for array shape reconstruction through mutual coupling," *IEEE Trans. Microw. Theory Techn.*, vol. 69, no. 10, pp. 4422–4436, Oct. 2021.
[34] L. H. Hsieh et al., "Development of a retrodirective wireless microwave power transmission system," in *Proc. IEEE Antennas Propag. Soc. Int. Symp.*, Columbus, OH, USA, 2003, pp. 393–396.
[35] X. Wang, B. Ruan, and M. Lu, "Retro-directive beamforming versus retro-reflective beamforming with applications in wireless power transmission," *Prog. Electromagn. Res.*, vol. 157, pp. 79–91, May 2016.
[36] E. Zaitsev and J. Hoffman, "Phased array flatness effects on antenna system performance," in *Proc. IEEE Int. Symp. Phased Array Syst. Technol.*, Waltham, MA, USA, 2010, pp. 121–125.
[37] C. Wang, M. Kang, W. Wang, B. Duan, L. Lin, and L. Ping, "On the performance of array antennas with mechanical distortion errors considering element numbers," *Int. J. Electron.*, vol. 104, no. 3, pp. 462–484, 2017.
[38] H. S. C. Wang, "Performance of phased array antennas with mechanical errors," in *Proc. IEEE Conf. Aerosp. Appl.*, Vail, CO, USA, 1990, pp. 317–318.
[39] N. J. Currie-Gregg and B. Peacock, "International space station robotic systems operations—A human factors perspective," *Proc. Hum. Factors Ergonom. Soc. Annu. Meeting*, vol. 46, no. 1, pp. 26–30, Sep. 2002.

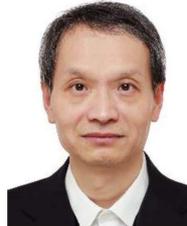

**Xinbin Hou** received the B.S. degree in mechanical engineering from North China Electric Power University, Beijing, China, in 1993, the M.S. degree in electro-mechanical Engineering from Xidian University, Xi'an, China, in 1998, and the Ph.D degree in man-machine and environmental engineering from Beihang University, Beijing, China, in 2002.

He is a Senior Researcher with the Qian Xuesen Laboratory of Space Technology, China Academy of Space Technology, Beijing, China. He has been engaged in solar power satellite research since 2006. He is currently the Head of SPS Research Project in China. He is the Secretariat of IAF Space Power Committee, the member of IAA Permanent Committee on Space Solar Power, and the member and Vice Secretary General of Space Solar Power Committee, Chinese Society of Astronautics. He has authored or coauthored four books (including two translations) and applied over 10 patents and more than 50 papers special in the field of SPS.

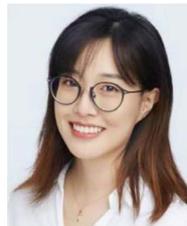

**Zhengai Cheng** received the B.S. degree in aircraft design and engineering from the Nanjing University of Aeronautics and Astronautics, Nanjing, China, in 2013.

She is currently a senior engineer at Qian Xuesen Laboratory of Space Technology, China Academy of Space Technology, Beijing, China. She is a member of Space Solar Power Committee, Chinese society of astronautics. Her research interests include in-space autonomous robotic assembly of space solar power station, deployable structure design, and ultra-high-speed collision simulation.

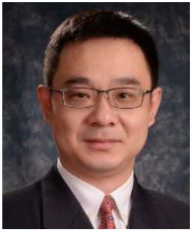

**Xin Wang** (Member, IEEE) received the B.S. and M.S. degrees in electronic engineering from Tsinghua University, Beijing, China, in 2000 and 2002, respectively, and the Ph.D. degree in electrical and computer engineering from Purdue University, West Lafayette, IN, USA, in 2009.

He was a Postdoctoral Research Associate with Birck Nanotechnology Center, Purdue University, until 2010. From 2010 to 2021, he was an Associate Professor with the College of Electronic and Information Engineering, Nanjing University of Aeronautics and Astronautics. He is currently a Professor with the School of Electrical Engineering, Chongqing University, Chongqing, China. His research interests include microwave power transmission, reconfigurable antennas, and RF circuits.

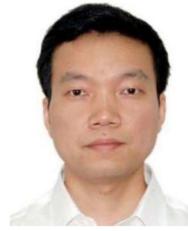

**Changjun Liu** (Senior Member, IEEE) received the B.S. degree in applied physics from Hebei University, Baoding, China, in 1994, and the M.S. and Ph.D. degrees in radio physics and biomedical engineering from Sichuan University, Chengdu, China, in 1997 and 2000, respectively.

From 2000 to 2001, he was a Postdoctoral Researcher with Seoul National University, Seoul, South Korea. From 2006 to 2007, he was a Visiting Scholar with Ulm University, Ulm, Germany. Since 1997, he has been with the Department of Radio Electronics, Sichuan University, where he has been a Professor since 2004. He has authored 2 books and more than 100 articles. His current research interests include microwave wireless power transmission and microwave power industrial applications.

Dr. Liu was a recipient of several honors, such as the outstanding reviewer for the IEEE TRANSACTIONS ON MICROWAVE THEORY AND TECHNIQUES, from 2006 to 2010, support from the MOE under the Program for New Century Excellent Talents in the University, China, from 2012 to 2014, and the Sichuan Province Academic technology leader since 2023.